\documentclass[3p, twocolumn, 10pt]{elsarticle}
\usepackage[T1]{fontenc}
\usepackage[english]{babel}
\biboptions{sort&compress}

\usepackage{hyperref}

\usepackage{amsmath}
\usepackage{amsfonts,amssymb}
\usepackage{multicol}

\usepackage{booktabs}
\usepackage[T1]{fontenc}
\usepackage{bm}
\usepackage{cuted}
\usepackage{graphicx}
\usepackage{lscape}
\usepackage{subfig}
\usepackage{float,caption}
\usepackage{mathtools}
\usepackage{fontawesome5}

\makeatletter
\def\ps@pprintTitle{%
  \let\@oddhead\@empty
  \let\@evenhead\@empty
  \def\@oddfoot{\reset@font\hfil\thepage\hfil}
  \let\@evenfoot\@oddfoot
}
\newcommand*{\smallrel}[2][.8]{%
  \mathrel{\mathpalette{\smallrel@{#1}}{#2}}%
}
\newcommand*{\smallrel@}[3]{%
  \sbox0{$#2\vcenter{}$}%
  \dimen@=\ht0 %
  \raise\dimen@\hbox{%
    \scalebox{#1}{%
      \raise-\dimen@\hbox{$#2#3\m@th$}%
    }%
  }%
}
\makeatother

\def\({\left(}
\def\){\right)}

\newcommand\rmd { {\rm d} }

\usepackage{tikz}
\usepackage{slashed}
\usepackage{pgfplots}
\usepackage{mathtools}
\usepackage{siunitx}
\usepackage{physics}
\usepackage[compat=1.1.0]{tikz-feynman}
\tikzset{graviton/.style={decorate, decoration={snake, amplitude=.6mm, segment length=1.5mm, pre length=.3mm, post length=.3mm}, double}}

\usepackage{xcolor}
\xdefinecolor{redviolet}{RGB}{228,38,207}
\definecolor{pastelgreen}{RGB}{210, 245, 210}
\definecolor{2gFSR}{RGB}{117,20,124}
\definecolor{2gFSRlight}{RGB}{210,179,213}
\definecolor{2gISR}{RGB}{55,126,33}
\definecolor{2gISRlight}{RGB}{185,215,181}
\definecolor{1gISR}{RGB}{202,102,39}
\definecolor{1gISRlight}{RGB}{237,208,182}
\definecolor{blueblob}{HTML}{4C72B0}
\xdefinecolor{brickred}{RGB}{220,60,66}
\xdefinecolor{bluscuro}{RGB}{51,51,179}

\usepackage{amssymb}

\begin{document}

\begin{frontmatter}
\title{\textbf{Structure-dependent radiative corrections to $e^+ e^- \to \pi^+ \pi^- \gamma$\\ in the GVMD approach}}

\author[a]{Carlo M. Carloni Calame}
\ead{carlo.carloni.calame@pv.infn.it}

\author[b,a]{Marco Ghilardi}
\ead{marco.ghilardi01@universitadipavia.it}

\author[c,d]{Andrea Gurgone}
\ead{andrea.gurgone@df.unipi.it}

\author[b,a]{Guido Montagna}
\ead{guido.montagna@unipv.it}

\author[e,f]{Mauro Moretti}
\ead{mauro.moretti@unife.it}

\author[a]{Oreste Nicrosini}
\ead{oreste.nicrosini@pv.infn.it}

\author[a,g]{Fulvio Piccinini}
\ead{fulvio.piccinini@pv.infn.it}

\author[b,a]{Francesco P. Ucci}
\ead{francesco.ucci@pv.infn.it}

\affiliation[a]{organization={INFN, Sezione di Pavia},
            addressline={Via A. Bassi 6},
            postcode={27100},
            city={Pavia},
            country={Italy}}
\affiliation[b]{organization={Dipartimento di Fisica ``Alessandro Volta'', \unexpanded{Universit\`a} di Pavia},
            addressline={Via A. Bassi 6},
            postcode={27100},
            city={Pavia},
            country={Italy}}
\affiliation[c]{organization={Dipartimento di Fisica, \unexpanded{Universit\`a} di Pisa},
            addressline={Largo Bruno Pontecorvo 3},
            postcode={56127},
            city={Pisa},
            country={Italy}}
\affiliation[d]{organization={INFN, Sezione di Pisa},
            addressline={Largo Bruno Pontecorvo 3},
            postcode={56127},
            city={Pisa},
            country={Italy}}
\affiliation[e]{organization={Dipartimento di Fisica e Scienze della Terra, \unexpanded{Universit\`a} di Ferrara},
            addressline={Via Saragat 1},
            postcode={44122},
            city={Ferrara},
            country={Italy}}
\affiliation[f]{organization={INFN, Sezione di Ferrara},
            addressline={Via Saragat 1},
            postcode={44122},
            city={Ferrara},
            country={Italy}}
\affiliation[g]{organization={INFN, Galileo Galilei Institute for 
Theoretical Physics},
            addressline={Largo E. Fermi 2},
            postcode={50125},
            city={Firenze},
            country={Italy}}

\begin{abstract}
We compute the radiative corrections to the process of two-pion production in association with a hard photon in 
$e^+ e^-$ annihilation by taking into account the non-perturbative structure of the pion in the one-loop calculation. 
For this purpose, we adopt the generalised vector meson dominance model to insert the pion form factor in loop integrals for the treatment of final-state radiation and initial-final state interference at next-to-leading order. 
We compare our predictions with the results
of the naive factorised scalar QED approach for experimentally relevant observables in the measurement of the  $e^+ e^- \to \pi^+ \pi^- \gamma$ process. 
The computation
 extends previous results obtained for the energy scan process $e^+ e^- \to \pi^+ \pi^-$
 and can be used to quantify the uncertainty 
due to the model describing the pion-photon interaction in radiative return experiments at flavour factories.
\end{abstract}

\end{frontmatter}

\section{Introduction}
\label{sec:introduction}

The electromagnetic pion form factor $F_\pi(q^2)$ provides information about the non-perturbative structure of the pion, describing the pion-photon interaction beyond the point-like approximation.

For time-like momenta, the pion form factor has a 
non-trivial shape around the $\rho$-meson resonance and
it is a crucial ingredient for the 
data-driven dispersive computation~\cite{Jegerlehner:2017gek,Colangelo:2018mtw,Keshavarzi:2019abf,Benayoun:2019zwh,Davier:2019can,Aoyama:2020ynm,Keshavarzi:2024wow,Aliberti:2025beg} of the leading-order hadronic 
vacuum polarisation~(HVP) correction to the anomalous 
magnetic moment of the muon, ${a_\mu = (g - 2)_\mu / 2}$. It is extracted at low-energy $e^+ e^-$ colliders (flavour 
factories) in energy scan~\cite{CMD-3:2023alj,CMD-3:2023rfe,Achasov:2006vp,SND:2020nwa,CMD-2:2001ski,CMD-2:2005mvb,CMD-2:2006gxt,1989321}
and radiative return~\cite{BaBar:2012bdw,BESIII:2015equ,KLOE:2004lnj,KLOE:2008fmq,KLOE:2010qei,KLOE:2012anl,KLOE-2:2016mgi,KLOE-2:2017fda} experiments. The present measurements 
of $F_\pi (q^2)$ display discrepancies that significantly affect the 
data-driven evaluations of the HVP
contribution to $a_\mu$, which disagree with the lattice QCD calculations, see~\cite{Aliberti:2025beg} for a complete discussion. Moreover, the resulting prediction of $a_\mu$ based on the data-driven method is not compatible with the experimental world average~\cite{Muong-2:2025xyk}, except for the CMD-3 determination. This situation, which is dubbed the ``new muon $g-2$ puzzle'', requires a deeper understanding of the tensions between 
the pion form factor determinations by different experiments.

New measurements of $F_\pi (q^2)$ with sub-percent precision are expected in the near future, by using 
both the energy scan and the radiative return method~\cite{SND_orsay2025,BELLE_orsay2025,KLOE_orsay2025,BES3_orsay2025,Zhang:2026jvb,Polat:2026ysh}. To support these experimental efforts, 
precise theoretical predictions are highly demanded. In this letter, we focus on radiative return
measurements with the aim of providing state-of-the-art predictions for the process 
$e^+ e^- \to \pi^+ \pi^- \gamma$ in relation to the existing literature.

At flavour factories, the standard tool used in radiative 
return experiments is the event generator \textsc{Phokhara}~\cite{Rodrigo:2001jr,Rodrigo:2001kf,Kuhn:2002xg,Czyz:2002np,Czyz:2003ue,Campanario:2019mjh}, which includes
the full set of next-to-leading~(NLO) corrections
 to ${e^+ e^- \to \pi^+ \pi^- \gamma}$. 
More recently, the NLO corrections to the same process have been matched to a fully-exclusive Parton Shower (PS) in the event generator \textsc{BabaYaga@NLO}~\cite{Budassi:2026lmr}. 
 In both codes, 
the pion-photon interaction in the computation of final-state radiation~(FSR) 
and initial-final state interference~(IFI) is treated according to the factorised scalar QED (F$\times$sQED) approach\footnote{While completing this work, a calculation of a subset of the structure-dependent corrections to radiative pion pair production was presented in~\cite{PetitRosas:2026iuq} and interfaced to \textsc{Phokhara}. One-photon-exchange topologies ($1\gamma^*$) shown in Fig.~\ref{fig:topologies_gvmd} are not included in~\cite{PetitRosas:2026iuq}.}.
In this model, 
the point-like scalar QED amplitudes are multiplied by an overall pion form factor, which is evaluated at an appropriate virtuality ensuring the cancellation of infrared (IR) divergences~\cite{Tracz:2018nkj}. Therefore, the composite structure of the pion is neglected in the calculation of NLO virtual correction beyond the IR limit.

% \textcolor{blue}{
% In this model, 
% the point-like scalar QED amplitudes are multiplied by an overall pion form factor, which is evaluated at an appropriate invariant mass ensuring the cancellation of all soft divergences~\cite{Tracz:2018nkj}.
% Hence, outside the soft limit, the composite structure of the pion is neglected in the
% calculation of NLO virtual corrections.
% More recently, the NLO corrections to $e^+ e^- \to \pi^+ \pi^- \gamma$ in the F$\times$sQED approach have been matched to a fully-exclusive parton shower~(PS) in~\cite{Budassi:2026lmr}.}

However, it was recently emphasised that the insertion of the non-perturbative
structure of the pion in one-loop computations is essential to obtain sensible 
predictions for various observables 
of experimental interest~\cite{Ignatov:2022iou,Colangelo:2022lzg,Budassi:2024whw,Gurgone:2025cci,Fang:2025mhn,Colangelo:2025ivq,Colangelo:2025iad,Monnard:2021pvm,Flores-Baez:2025nra}. In particular, it was shown in~\cite{Ignatov:2022iou,Colangelo:2022lzg,Budassi:2024whw,Gurgone:2025cci,Fang:2025mhn} that the structure-dependent corrections 
beyond the point-like approximation are crucial for a 
correct description of the forward-backward (or charge) asymmetry of the
energy scan process $e^+ e^- \to \pi^+ \pi^-$.

In the literature, two methods are available to take into account the internal structure of the pion in the calculation of loop integrals:
the generalised vector meson dominance~(GVMD) model~\cite{Ignatov:2022iou,Budassi:2024whw} and 
the so-called FsQED approach~\cite{Colangelo:2022lzg,Budassi:2024whw,Gurgone:2025cci,Colangelo:2025ivq,Fang:2025mhn,Monnard:2021pvm}, based on dispersion relations. In this paper, we make use of the former strategy 
and apply it to the radiative process $e^+ e^- \to \pi^+ \pi^-\gamma$. The application of the dispersive approach to the same process is left to a future publication.

One of the main goals of this work is to compare the GVMD predictions with the 
results of the F$\times$sQED description in order to estimate
the uncertainty associated with the modelling of the pion-photon
interaction in the computation of the radiative corrections 
to $e^+ e^- \to \pi^+ \pi^-\gamma$ at NLO. This uncertainty 
is a source of systematic error in radiative 
return experiments and can now be assessed in a sound way thanks to recent 
theoretical progress in the field. 
Moreover, our results represent 
a direct improvement of the existing F$\times$sQED calculations. 
In this respect, our study extends the predictions 
for the structure-dependent corrections 
obtained in~\cite{Ignatov:2022iou,Colangelo:2022lzg,Budassi:2024whw,Gurgone:2025cci,Fang:2025mhn} from energy scan observables to 
radiative return measurements.

The contributions to $e^+ e^- \to \pi^+ \pi^-\gamma$ due to radiative meson decays around the $\phi$ resonance and other relevant processes have been studied in~\cite{Achasov:1997gb,Dubinsky:2004xv,Isidori:2006we,Pancheri:2007xt,Gallegos:2009qu,Roca:2009zy}. 
%The non-point-like interaction of pions to photons in the $e^+ e^- \to \pi^+ \pi^-\gamma$ process has been studied in~\cite{Achasov:1997gb,Dubinsky:2004xv,Isidori:2006we,Pancheri:2007xt,Gallegos:2009qu,Roca:2009zy}, with particular attention to the radiative meson decay processes around the $\phi$ resonance. 
These contributions, which are important for the KLOE experiment at DA$\Phi$NE, are not considered in the present work.

The structure of the article is as follows. In Section~\ref{sec:calculation},
we describe the details of the calculation of the one-loop 
corrections according to the GVMD model. In Section~\ref{sec:numres},
we show numerical results for experimentally relevant differential cross sections 
by using realistic
event selection criteria, inspired by radiative return experiments at flavour factories. In Section~\ref{sec:cons_checks}, we show some consistency checks of our GVMD implementation.
Section~\ref{sec:conclusion} summarises the main conclusions of 
our study.

\section{One-loop calculation in GVMD approach}
\label{sec:calculation}

In this section, we sketch the main features of the FSR and IFI one-loop corrections to  the process $e^+ e^- \to \pi^+ \pi^-\gamma$ according to the GVMD approach, inspired by the model of the same name~\cite{Sakurai:1972wk}. Since $F_\pi(0)=1$, the real corrections in the GVMD approach are the same as in F$\times$sQED. More details about this method can be found in~\cite{Ignatov:2022iou,Budassi:2024whw}.

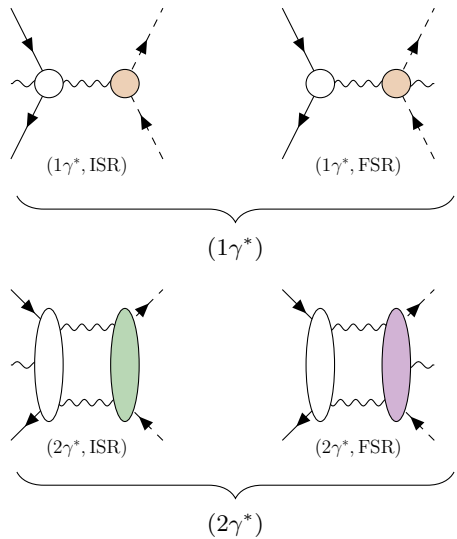
\begin{figure}
\vspace{-5mm}
\begin{minipage}{0.05\textwidth}
\,\,
\end{minipage}
% Seconda riga: ultimi due diagrammi
\begin{minipage}{0.2\textwidth}
\begin{tikzpicture}[scale=0.5, transform shape]
\begin{feynman}[baseline=(a.b),small]
    % External vertices
    \vertex (a) at (-2, -2) ;
    \vertex (b) at (-1, -1) ;
    \vertex (u) at (-1, 0) ;
    \vertex (p) at (-2, 0) ;
    \vertex (c) at (-1, 1) ;
    \vertex (d) at (-2, 2) ;
    \vertex [dot] (e) at (2, -2) ;
    \vertex (f) at (1, -1) ;
    \vertex (g) at (1, 1) ;
    \vertex [dot] (h) at (2, 2) ;
    \vertex (y) at (0, -2.2) {{\Large{\text{$(2\gamma^*,{\rm ISR})$}}}};

    % Connections
    \diagram* {
      (d) -- [fermion] (c) -- [fermion] (u) -- [fermion] (b) -- [fermion] (a),
      (e) -- [charged scalar] (f) -- [charged scalar] (g) -- [charged scalar] (h),
      (g) -- [photon] (c),
      (b) -- [photon] (f),
      (p) -- [photon] (u),
    };
\end{feynman}

\begin{scope}
\node[draw=black, ellipse, minimum width=0.75cm, minimum height=3cm, xshift=-1cm, fill=white] {};
\node[draw=black, ellipse, minimum width=0.75cm, minimum height=3cm, xshift=1cm, fill=2gISRlight] {};
% \node[draw=black, ellipse, minimum width=0.75cm, minimum height=3cm, xshift=1cm, fill=black, pattern=north west lines, pattern color=black] {};
\end{scope}
\end{tikzpicture}
\end{minipage}%
\hspace{0.01\textwidth}
\begin{minipage}{0.2\textwidth}
\begin{tikzpicture}[scale=0.5, transform shape]
\begin{feynman}[baseline=(a.b),small]
    % External vertices
    \vertex (a) at (-2, -2) ;
    \vertex (b) at (-1, -1) ;
    \vertex (u) at (1, 0) ;
    \vertex (p) at (2, 0) ;
    \vertex (c) at (-1, 1) ;
    \vertex (d) at (-2, 2) ;
    \vertex [dot] (e) at (2, -2) ;
    \vertex (f) at (1, -1) ;
    \vertex (g) at (1, 1) ;
    \vertex [dot] (h) at (2, 2) ;
    \vertex (y) at (0, -2.2) ;
    \vertex (y) at (0, -2.2) {\Large{\text{$(2\gamma^*,{\rm FSR})$}}};
    
    % Connections
    \diagram* {
      (d) -- [fermion] (c)-- [fermion] (b) -- [fermion] (a),
      (e) -- [charged scalar] (f) -- [charged scalar] (g) -- [charged scalar] (h),
      (g) -- [photon] (c),
      (b) -- [photon] (f),
      (p) -- [photon] (u),
    };
\end{feynman}

\begin{scope}
\node[draw=black, ellipse, minimum width=0.75cm, minimum height=3cm, xshift=1cm, fill=2gFSRlight] {};
% \node[draw=black, ellipse, minimum width=0.75cm, minimum height=3cm, xshift=1cm, fill=black, pattern=north west lines, pattern color=black] {};
\node[draw=black, ellipse, minimum width=0.75cm, minimum height=3cm, xshift=-1cm, fill=white] {};
\end{scope}
\end{tikzpicture}
\end{minipage}%
\begin{center}
\vspace{-3mm}
\begin{tikzpicture}
\node (A) at (0,0) {};
\node (B) at (6,0) {}; % regola la lunghezza della graffa

\draw[decorate, decoration={brace, mirror, amplitude=10pt}]
(A) -- (B)
node[midway, below=12pt] {{\small{\text{$(2\gamma^*)$}}}};
\end{tikzpicture}
\end{center}
\vspace{-4mm}
\caption{Topologies involved in the calculation of 
the structure-dependent corrections to $e^+e^-\to\pi^+\pi^-\gamma$. $1\gamma^*$ and $2\gamma^*$ refer to the number of virtual photons connecting initial-state and final-state charged legs, while ISR and FSR are related to the signal photon.}
\label{fig:topologies_gvmd}
\end{figure}

When considering $e^+e^-\to\pi^+\pi^-\gamma$ at NLO, the four topologies shown in Fig.~\ref{fig:topologies_gvmd} contribute. In a general calculation, both the white and coloured blobs include 
all possible radiative corrections
and additional higher-order tensor structures appear.
%(andando su con gli ordini radiativi ci sono anche scmabi di n fotoni, nonche' correzioni di QCD low energy anche per la blob leptonica) }. 
However, in the 
present work, we perform a genuine NLO calculation and, 
therefore, each topology should be understood as the sum of all contributions up to $\mathcal{O}(\alpha)$. The white blob can be computed using perturbative QED, since the particles involved are leptons. In contrast, the coloured blob contains non-perturbative effects due to the presence of the pion form factor. Hence, it requires a more careful treatment.

As discussed in~\cite{Aliberti:2024fpq}, the computation of the structure-dependent corrections to radiative return involves the matrix elements associated to the sub-processes $\gamma^* \gamma^* \to \pi^+\pi^-$  ($2\gamma^*,{\rm ISR}$), $i.e.$ the pion Compton tensor, and $\gamma^* \gamma^* \to \pi^+\pi^-\gamma$ ($2\gamma^*,{\rm FSR}$). Moreover, also corrections to the pion form factor ($1\gamma^*,{\rm ISR}$) and to  
the Compton tensor ($1\gamma^*,{\rm FSR}$) come into play. 
The $\gamma^* \gamma^* \to \pi^+\pi^-$ and $\gamma^* \to \pi^+\pi^-$ contributions are already present in the calculation of the radiative corrections to the two-pion production, for which the GVMD model is in good agreement with the dispersive method~\cite{Colangelo:2022lzg,Budassi:2024whw}. 
Therefore, for the ($1\gamma^*,{\rm ISR}$) topology, we can a priori expect that the GVMD approach has the same level of accuracy as in the $e^+ e^- \to \pi^+ \pi^-$ case. On the other hand, the reliability of the GVMD model to the ($2\gamma^*,{\rm ISR}$) subset can be affected by the presence of an additional momentum scale in the loop integral related to pentagon diagrams, that could enhance unphysical sub-threshold imaginary parts. For such contributions, we pragmatically assume that the GVMD is a good approximation. For radiative corrections to $\gamma^* \to \pi^+\pi^-\gamma$ and for the subprocess $\gamma^* \gamma^* \to \pi^+\pi^-\gamma$, dispersive analyses are still in progress~\cite{Aliberti:2024fpq,Hoferichter_Liverpool2025,Kaziukenas_Liverpool2025}. Also for these topologies, we make the assumption that GVMD is an appropriate framework to compute the structure-dependent corrections to radiative two-pion production.

\begin{figure}[t]
\begin{minipage}{0.05\textwidth}
\,\,
\end{minipage}
% Prima riga: primi due diagrammi
\begin{minipage}{0.2\textwidth}
\begin{tikzpicture}[scale=0.5, transform shape]
\begin{feynman}[baseline=(a.b),small]
    % External vertices
    \vertex (a) at (-2, -2) ;
    \vertex (b) at (-1, -1) ;
    \vertex (u) at (-1, 0) ;
    \vertex (p) at (-2, 0) ;
    \vertex (c) at (-1, 1) ;
    \vertex (d) at (-2, 2) ;
    \vertex [dot] (e) at (2, -2) ;
    \vertex (f) at (1, -1) ;
    \vertex (g) at (1, 0) ;
    \vertex [dot] (h) at (2, 2) ;
    \vertex (y) at (0, 2) ;
    \vertex (r) at (0, -2.2) {{\Large{\text{$(1\gamma^*,{\rm ISR})$}}}};
    % Connections
    \diagram* {
      (d) -- [fermion]  (u) -- [fermion] (a),
      (e) -- [charged scalar] (g) -- [charged scalar] (h),
      (g) -- [photon] (u),
      (p) -- [photon]  (u),
    };
\end{feynman}

\begin{scope}
\node[draw=black, circle, fill=white, minimum width=0.75cm, minimum height=0.75cm, xshift=-1cm] {};
\node[draw=black, circle, fill=1gISRlight, minimum width=0.75cm, minimum height=0.75cm, xshift=1cm] {};
% \node[draw=black, circle, fill=white, minimum width=0.75cm, minimum height=0.75cm, xshift=1cm, pattern=north west lines, pattern color=black] {};
\end{scope}
\end{tikzpicture}
\end{minipage}%
\hspace{0.01\textwidth}
\begin{minipage}{0.2\textwidth}
\begin{tikzpicture}[scale=0.5, transform shape]
\begin{feynman}[baseline=(a.b),small]
    % External vertices
    \vertex (a) at (-2, -2) ;
    \vertex (b) at (-1, -1) ;
    \vertex (u) at (-1, 0) ;
    \vertex (p) at (-2, 0) ;
    \vertex (c) at (-1, 1) ;
    \vertex (d) at (-2, 2) ;
    \vertex [dot] (e) at (2, -2) ;
    \vertex (f) at (1, -1) ;
    \vertex (g) at (1, 0) ;
    \vertex [dot] (h) at (2, 2) ;
    \vertex (y) at (2,0) ;
    \vertex (r) at (0, -2.2) {{\Large{\text{$(1\gamma^*,{\rm FSR})$}}}};
    
    % Connections
    \diagram* {
      (d) -- [fermion]  (u) -- [fermion] (a),
      (e) -- [charged scalar] (g) -- [charged scalar] (h),
      (g) -- [photon] (u),
      (y) -- [photon]  (g),
    };
\end{feynman}

\begin{scope}
\node[draw=black, circle, fill=white, minimum width=0.75cm, minimum height=0.75cm, xshift=-1cm] {};
\node[draw=black, circle, fill=1gISRlight, minimum width=0.75cm, minimum height=0.75cm, xshift=1cm] {};
% \node[draw=black, circle, fill=white, minimum width=0.75cm, minimum height=0.75cm, xshift=1cm, pattern=north west lines, pattern color=black] {};
\end{scope}
\end{tikzpicture}
\end{minipage}%

\begin{center}
\vspace{-3mm}
\begin{tikzpicture}
\node (A) at (0,0) {};
\node (B) at (6,0) {}; % regola la lunghezza della graffa

\draw[decorate, decoration={brace, mirror, amplitude=10pt}]
(A) -- (B)
node[midway, below=12pt] {{\small{\text{$(1\gamma^*)$}}}};
\end{tikzpicture}
\end{center}

\end{figure}

In the GVMD approach, the pion form factor is approximated to a sum of a finite number $n_r$ of Breit-Wigner (BW) functions, and it is given by:
\begin{align}
F^\text{BW}_\pi(q^2) = \sum_{v=1}^{n_r} F^\text{BW}_{\pi,v}(q^2)  = \frac{1}{c_t} \sum_{v=1}^{n_r} c_v \frac{\Lambda_v^2}{\Lambda_v^2 - q^2} \:,
\label{eq:bwsum}
\end{align}
where $\Lambda_v^2 = m_v^2 - i m_v \Gamma_v$ and $c_v = |c_v|e^{i\phi_v}$. The division by $c_t = \sum_v c_v$ 
ensures the normalisation condition $F^\text{BW}_\pi(0)=1$.
In the GVMD model, each form factor term $F^\text{BW}_{\pi,v}(q^2)$ corresponds to the propagator of a vector meson $v$ with 
mass $m_v$ and width $\Gamma_v$, multiplied by a complex coupling $c_v$.
As discussed in~\cite{Ignatov:2022iou}, the pion form factor is inserted in the amplitudes by modifying each sQED vertex
according to the following 
rules:\\
\begin{align*}
%\,\\[-2.5\baselineskip]
\;
\begin{gathered}
\begin{tikzpicture}[baseline=(a)]
\begin{feynman}[inline=(a)]
    \vertex (a);
    \vertex[right=0.5cm of a,style=dot] (b) {};
    \vertex[right = 0.5cm of b, style=dot] (e) {};
    \vertex[above right=0.75cm and 0.75cm of e] (c);
    \vertex[below right=0.75cm and 0.75cm of e] (d);
    \diagram* {
      (a) -- [photon] (b),
      (b) --[graviton] (e),
      (c) -- [scalar] (e),
      (e) -- [scalar] (d),
      (a) -- [fermion,opacity=0.0] (e)
    };
\end{feynman}
\end{tikzpicture}
\end{gathered}
\,\,&=\;\;\;
\begin{gathered}
\begin{tikzpicture}[baseline=(a)]
\begin{feynman}[inline=(a)]
    \vertex (a);
    \vertex[right=1.0cm of a] (b);
    \vertex[above right=0.75cm and 0.75cm of b] (c);
    \vertex[below right=0.75cm and 0.75cm of b] (d);
    \diagram* {
      (a) -- [photon, momentum'=$q$] (b),
      (c) -- [scalar] (b),
      (b) -- [scalar] (d),
    };
\end{feynman}
\end{tikzpicture}
\end{gathered}
{\small \text{$\times\!$ $F^\text{BW}_\pi(q^2) $}}\,,
\\[2pt]
\;
\begin{gathered}
\begin{tikzpicture}[baseline=(a)]
\begin{feynman}[inline=(a)]
     \vertex[style=dot] (a) {};
      \vertex[above left=0.75cm and 0.75cm of a] (b);
     \vertex[below left=0.75cm and 0.75cm of a] (e);
       \vertex[above left=0.35cm and 0.35  cm of a, style=dot] (f) {};
     \vertex[below left=0.35cm and 0.35  cm of a, style=dot] (g){};
     \vertex[above right=0.75cm and 0.75cm of a] (c);
     \vertex[below right=0.75cm and 0.75cm of a] (d);
     \diagram* {
       (e) -- [photon] (g),
       (b) -- [photon] (f),
       (e) -- [photon] (g),
       (b) -- [photon] (f),
       (c) -- [scalar] (a),
       (a) -- [scalar] (d),
       (g) -- [graviton] (a),
       (f) -- [graviton] (a),
       (e) -- [photon, opacity=0.0] (a),
       (b) -- [photon, opacity=0.0] (a)
     };
\end{feynman}
\end{tikzpicture}
\end{gathered}
\,\, &=\,\,
\begin{gathered}
\begin{tikzpicture}[baseline=(a)]
\begin{feynman}[inline=(a)]
        \tikzfeynmanset{
    momentum/arrow shorten=0.25,
    }
     \vertex (a);
      \vertex[above left=0.75cm and 0.75cm of a] (b);
     \vertex[below left=0.75cm and 0.75cm of a] (e);
     \vertex[above right=0.75cm and 0.75cm of a] (c);
     \vertex[below right=0.75cm and 0.75cm of a] (d);
     \diagram* {
       (e) -- [photon, momentum'=$q_2$] (a),
       (b) -- [photon, momentum'=$q_1$] (a),
       (c) -- [scalar] (a),
       (a) -- [scalar] (d),
     };
\end{feynman}
\end{tikzpicture}
\end{gathered}
    {\small \text{$\times\!$ $ F^\text{BW}_\pi(q_1^2)\,F^\text{BW}_\pi(q_2^2) $}}\,.
\end{align*}
In this way, gauge invariance is preserved and, since each contribution $F^\text{BW}_{\pi,v}(q^2)$ has a propagator-like structure, the virtual amplitudes can be computed using standard loop techniques. 
As already pointed out in~\cite{Budassi:2024whw} for the  $e^+e^-\to\pi^+\pi^-$ case, proper analytical tricks can be adopted in such a way that the number of loop propagators does not increase with respect to the point-like calculation. 
For all the topologies shown in Fig.~\ref{fig:topologies_gvmd}, we can write the virtual one-loop amplitude as:
\begin{align}\label{eq:gvmd_fsr}
\begin{split}
    \mathcal{A}_{\text{GVMD}}= 2\Re{\sum_{v,w=1}^{n_r} \frac{c_v \hspace{1pt} c_w}{c_t^2}\,\mathcal{A}_{\text{GVMD}}(\Lambda_v^2,\Lambda_w^2)}
    \,,
\end{split}
\end{align}
where the $(1\gamma^*)$ and $(2\gamma^*)$ contributions can be explicitly written as in Eq.~\eqref{eq:fsrifi_gvmd}.
In this expression, indices $i,j$ stand for the current emitting the real photon in the virtual and Born matrix elements, respectively. The virtuality of the pion form factor $\eta$ is equal to $M_{\pi\pi}^2$ for ISR and $s$ for FSR, $M_{\pi\pi}$ being the di-pion invariant mass and $s$ the squared centre-of-mass (c.m.) energy.
\newpage
\begin{strip}
\noindent
% \rule{0.5\textwidth}{0.4pt}%
% \hspace{-0.4pt}%
% \rule{0.4pt}{6pt}
{\small \begin{equation}
\begin{split}
&\mathcal{A}_{\text{GVMD}}^{(1\gamma^*)}(\Lambda_v^2,\Lambda_w^2)=\sum_{i,j=\rm ISR,FSR}\left\{\mathcal{A}^{(1\gamma^*)}_{i,j} \left(\lambda^2 \right)+\frac{1}{\Lambda_v^2-\Lambda_w^2}\left[\Lambda_w^2\,\mathcal{A}^{(1\gamma^*)}_{i,j}\left(\Lambda_v^2\right)-\Lambda_v^2\,\mathcal{A}^{(1\gamma^*)}_{i,j}\,\left(\Lambda_w^2\right)\right]\right\}F_\pi(\eta_i)F^*_\pi(\eta_j)\,,\\
&\mathcal{A}_{\text{GVMD}}^{(2\gamma^*)}(\Lambda_v^2,\Lambda_w^2)=\sum_{i,j=\rm ISR,FSR}\left\{\mathcal{A}^{(2\gamma^*)}_{i,j} \left(\lambda^2 ,\lambda^2\right)-\mathcal{A}^{(2\gamma^*)}_{i,j} \left(\Lambda_v^2 ,\lambda^2\right)-\mathcal{A}^{(2\gamma^*)}_{i,j} \left(\lambda^2 ,\Lambda_w^2\right)+\mathcal{A}^{(2\gamma^*)}_{i,j} \left(\Lambda_v^2 ,\Lambda_w^2\right)\right\}F^*_\pi(\eta_j)\,.
\end{split}
\label{eq:fsrifi_gvmd}
\end{equation}}
\end{strip}

The quantities {\small$\mathcal{A}^{(1\gamma^*)}_{i,j}\!\left(\Lambda^2\right)$} represent the $(1\gamma^*)$ virtual point-like amplitude with the loop photon having complex squared mass $\Lambda^2$. In particular, when $v=w$, the derivative of {\small$\mathcal{A}^{(1\gamma^*)}_{i,j}\!\left(\Lambda_v^2\right)$}  with respect to the resonance complex mass $\Lambda_v^2$ must be evaluated and it is computed numerically in the present study. On the other hand, {\small$\mathcal{A}^{(2\gamma^*)}_{i,j}\!\left(\Lambda_1^2,\Lambda_2^2\right)$} is the $(2\gamma^*)$ virtual point-like amplitude where the photons involved in the loop have complex squared masses $\Lambda_1^2$ and $\Lambda_2^2$.

The fictitious photon mass $\lambda$ is used for the regularisation of IR singularities. 
By construction, in the GVMD approach, the IR behaviour of the virtual corrections is the same as the F$\times$sQED amplitudes, and we checked that our results are $\lambda$-independent by varying its value.

The computation is performed in dimensional regularisation and on-shell renormalisation scheme for the treatment of the ultraviolet divergences. The pion wave function counterterm can be found in~\cite{Budassi:2024whw}, while the mass counterterm is given by:\footnote{We omit tadpole contributions, as they do not contribute to the renormalised amplitude.} 
{ \begin{equation}
\begin{split}
    \delta Z_{m_\pi}\left(\Lambda_v^2\right)&=\frac{\alpha}{4\pi \,m_\pi^2}\Bigl\{{\rm A}_0\left(m_\pi^2\right)-2{\rm A}_0\left(\Lambda_v^2\right)
    \\
    &+(\Lambda^2_v-4m_\pi^2){\rm B}_0\left(m_\pi^2,\Lambda^2_v,m_\pi^2\right)\Bigr\}.
\end{split}
\end{equation}}

\begin{table*}[t]
\centering
\begin{tabular}{lcccccc}
\toprule
 & $\rho$  & $\omega$ & $\phi$& $\rho'$ & $\rho''$ & $\rho'''$ \\[1pt]
\midrule
$m_v$ (MeV) & 755.18 & 781.81 & 1019.47 & 1319.9 & 1732.2 & 2218.4 \\
$\Gamma_v$ (MeV) & 139.62& 7.1726 & 4.8770 & 698.94 & 320.02 & 123.50  \\
$|c_v|$ & - & 0.0075 & 0.00074 & 0.2395 & 0.0876 & 0.0064 \\
$\varphi_v$ (rad) & -& 1.67601  & 5.5833 &3.0948 &0.5447 & -0.2988  \\
\bottomrule
\end{tabular}
\caption{Input parameters for our model of the pion 
form factor $F^{\rm BW}_\pi(q^2)$.}
\label{tab:vff}
\end{table*}

The amplitudes with arbitrary resonance masses are generated using the \textsc{FeynArts}~\cite{Hahn:1998yk,Hahn:2000kx,Hahn:2010zi}$\to$\textsc{FeynCalc}~\cite{Shtabovenko:2016sxi,Shtabovenko:2020gxv,Shtabovenko:2023idz} chain and are expressed as functions of one-loop tensor integrals. The latter are reduced in terms of scalar integrals up to 4-points\footnote{The 5-point tensor integrals are written as sum of 4-point and 3-point scalar functions using the package \texttt{hexagon.m}~\cite{Diakonidis:2008ij}.}. The scalar loop functions are finally evaluated using the $\textsc{Collier}$~\cite{Denner:2014gla} library, with the exception of the numerical derivative of the amplitude, which is computed in quadruple precision using the \textsc{LoopTools}~\cite{Hahn:1998yk} package.

The above calculation has been cross-checked with an independent computation implemented in \textsc{Form}~\cite{Vermaseren:2000nd,Kuipers:2013pba,Ruijl:2017dtg} and using \textsc{LoopTools}~\cite{Hahn:1998yk} to evaluate loop functions. We found perfect agreement between the two methods.

The full analytic amplitudes derived in this work are planned to be made publicly available as part of a future public release of the \textsc{BabaYaga@NLO} code through the project repository on \href{https://github.com/cm-cc/BabaYagaNLO}{GitHub \faGithub}.

\section{Numerical results}\label{sec:numres}

\begin{figure*}[htbp]
    \centering
    \includegraphics[width=0.95\textwidth]{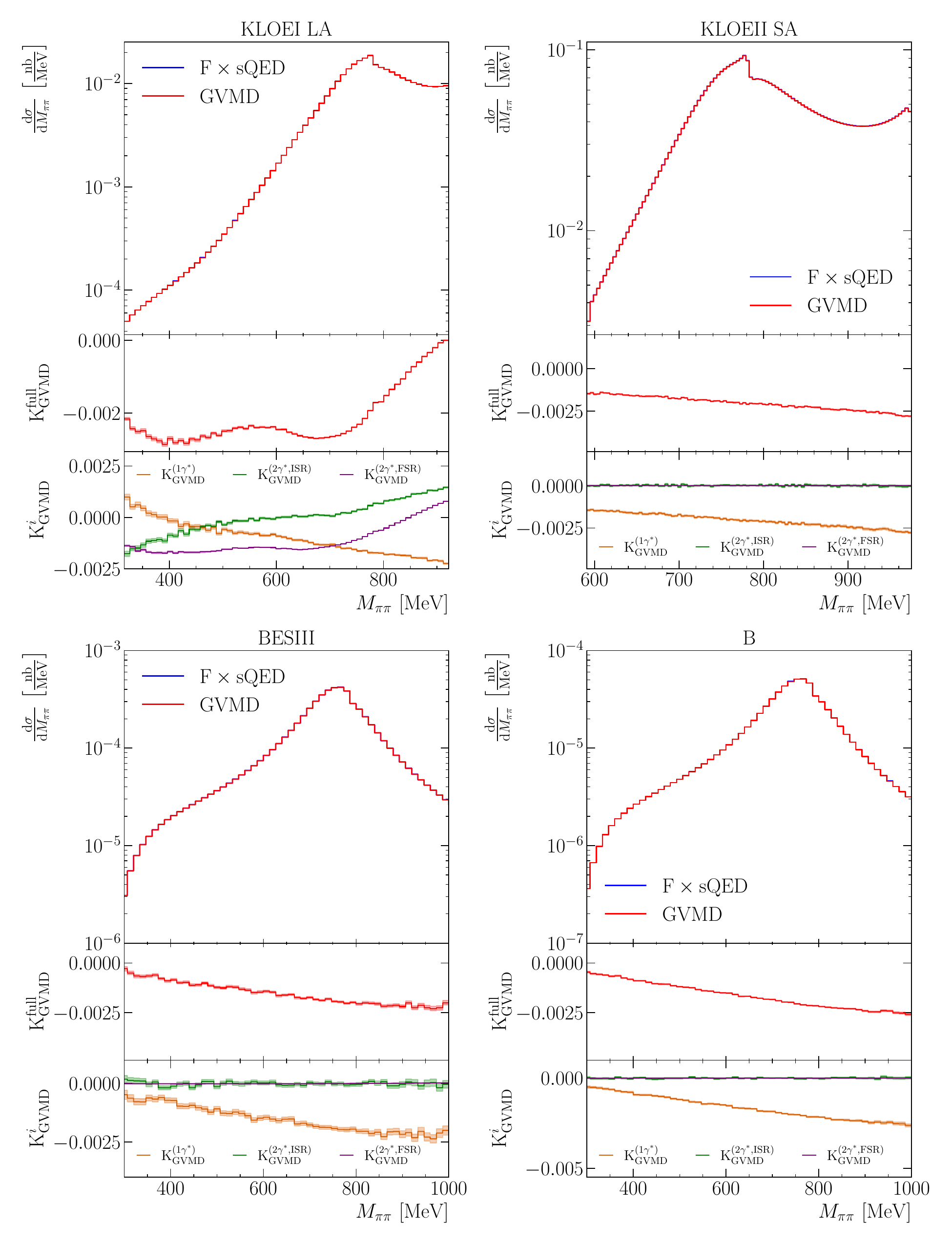}
    \caption{The differential cross section of the $e^+e^-\to\pi^+\pi^-\gamma$ process as a function of the invariant mass $M_{\pi\pi}$ in the four scenarios with the 
    F$\times$sQED and GVMD approaches. The ${\rm K}^i_{\rm GVMD}$ factors shown in the lower panels are defined in 
    Eq.~\eqref{eq:Kfactor}.}
    \label{fig:invmass}
    \end{figure*}

\begin{figure*}[htbp]
    \centering
    \includegraphics[width=0.95\textwidth]{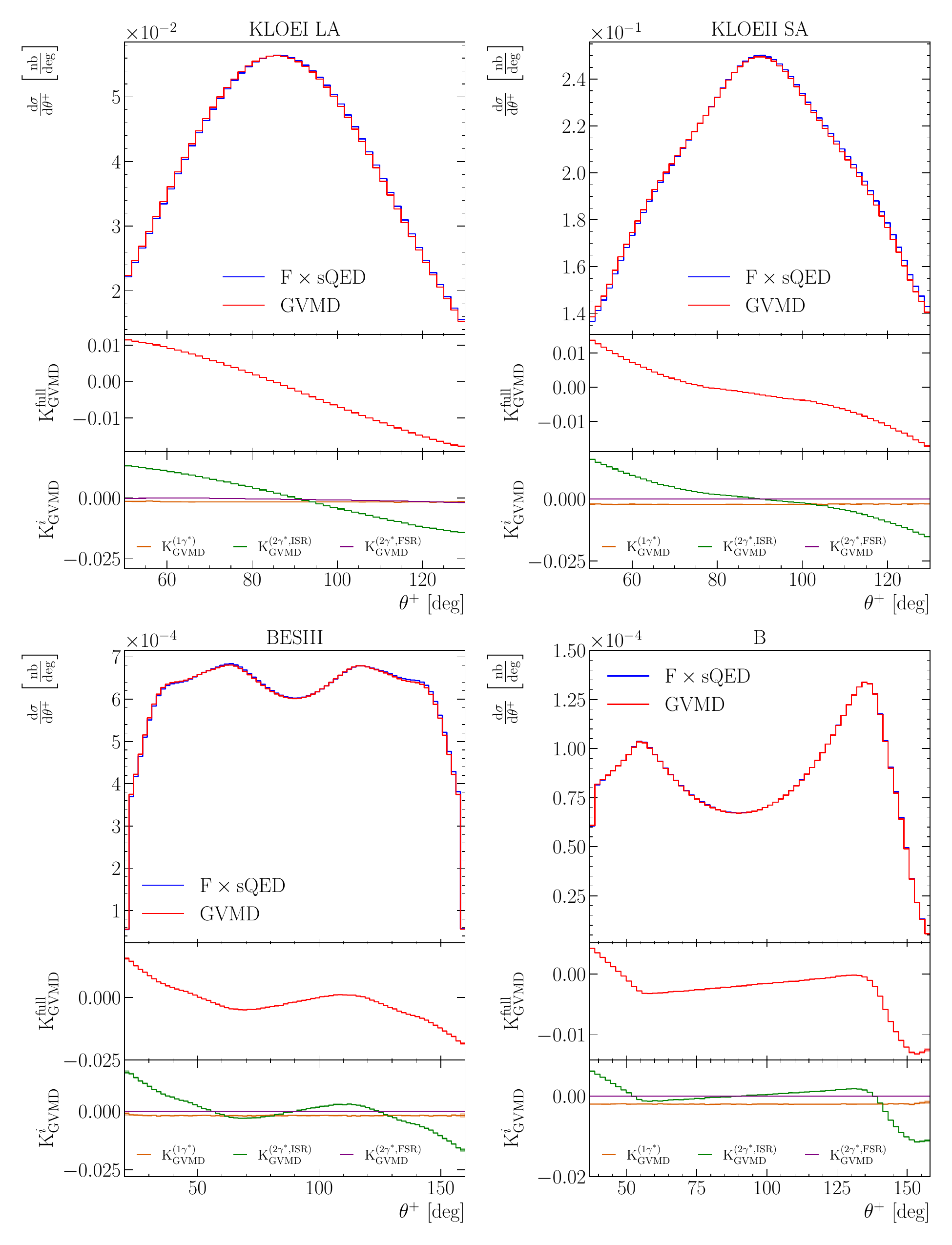}
    \caption{The same as in Fig.~\ref{fig:invmass} for the
    differential cross section as a function of the $\theta^+$ scattering angle.    
    }
    \label{fig:angdist}
\end{figure*}

We have implemented the NLO calculation described in Section~\ref{sec:calculation} in the Monte Carlo event generator \textsc{BabaYaga@NLO}~\cite{Budassi:2024whw,Budassi:2026lmr,Balossini:2006wc,Balossini:2008xr}. The GVMD fixed-order corrections have been matched with the PS algorithm reaching NLOPS accuracy, as done in~\cite{Budassi:2026lmr} for the factorised approach. 
Since the higher orders are modified by sub-leading terms with respect to the $\rm F \times sQED$ case, only the leading structure-dependent effects appearing at NLO are shown.
%Since the differences between the ${\rm F}\times {\rm sQED}$ and GVMD approaches arise already at NLO, for the sake of simplicity we present numerical results at this perturbative order.

For the numerical results, we use the following set of input parameters:
{\begin{equation}
    \begin{split}
    \alpha & = \, 1/137.03599908 \,, \\
     m_e & = \, 0.51099895~{\rm MeV}, \\
     m_{\pi}& =\, 139.57039~{\rm MeV}. 
\end{split}
\end{equation}}
In the same spirit of~\cite{Budassi:2024whw}, 
the parameters entering the pion form factor used in our simulations are fitted to the available data
% \footnote{We do not claim that $F_\pi^\textrm{BW}(q^2)$ is a proper extraction of the pion form factor. 
% It has to be understood as a fixed parametrisation inspired by real data and used to study the impact of radiative corrections in a realistic scenario. For this reason, we do not assign an error to the parameters nor analyse the agreement with the data.
% }
of 
BABAR~\cite{BaBar:2012bdw}, BESIII~\cite{BESIII:2015equ}, 
CMD-2~\cite{CMD-2:2001ski,CMD-2:2005mvb,CMD-2:2006gxt}, 
CMD-3~\cite{CMD-3:2023alj,CMD-3:2023rfe}, 
DM-2~\cite{1989321}, KLOE~\cite{KLOE:2004lnj,KLOE:2008fmq,KLOE:2010qei,KLOE:2012anl,KLOE-2:2016mgi,KLOE-2:2017fda}, and SND~\cite{Achasov:2006vp,SND:2020nwa} experiments. We use the BW expression given by Eq.~(\ref{eq:bwsum}), considering six resonances ($\rho$, $\omega$, $\phi$, $\rho'$, $\rho''$, $\rho'''$) and we use this parametrisation for both the F$\times$sQED and GVMD results shown in the following. 
We perform the fit by imposing the sum rule 
\begin{equation}
    \frac{1}{\pi} \, \int_{4m_\pi^2}^{\infty} \ 
    \frac{\rmd s^{\prime}}{s^{\prime}}  \, {\rm Im} \, F_{\pi}^{\rm BW} (s^{\prime}) \, = \, 1 \,,
\end{equation}
which we find satisfied at the $10^{-3}$ level. Despite this requirement is not necessary to fit the data, it allows a much better agreement between a GVMD-like form factor and more refined parametrisations, for both the real and imaginary parts.
The fitted values for the six resonances are shown in Tab.~\ref{tab:vff}.

We do not claim that $F_\pi^{\mathrm{BW}}(q^2)$ constitutes a proper extraction of the physical pion form factor. It should instead be understood as a fixed phenomenological parametrisation inspired by experimental data and employed solely as a necessary input to investigate the impact of structure-dependent corrections. Furthermore, the identification $F_\pi(q^2) \simeq F_\pi^{\mathrm{BW}}(q^2)$ does not merely affect the precision of the form factor description. Indeed, since $\mathrm{Im}\, F_\pi^{\mathrm{BW}}(q^2 < 4m_\pi^2) \neq 0$, the GVMD framework does not implement unitarity in a natural way. In addition, when expressing the form factor as a sum of Breit–Wigner contributions, the treatment of the $\rho$–$\omega$ interference is only approximately captured at a phenomenological level. A more consistent and formal treatment of the pion form factor is provided in terms of a dispersive parametrisation, where the form factor satisfies 
analyticity and unitarity constraints, as well as incorporates the $\rho$–$\omega$ mixing contribution in a more rigorous way, 
as discussed %in~\cite{Colangelo:2018mtw,OConnell:1995fwv,OConnell:1995nse,Czyz:2010hj}.}
in~\cite{Colangelo:2018mtw,DeTroconiz:2001rip,Colangelo:2003yw,Colangelo:2022prz,deTroconiz:2004yzs,Ananthanarayan:2013zua,Ananthanarayan:2016mns,Hoferichter:2016duk,Hanhart:2016pcd}.

\begin{figure}[t] % [t] for top of page
\centering
\includegraphics[width=0.5\textwidth]{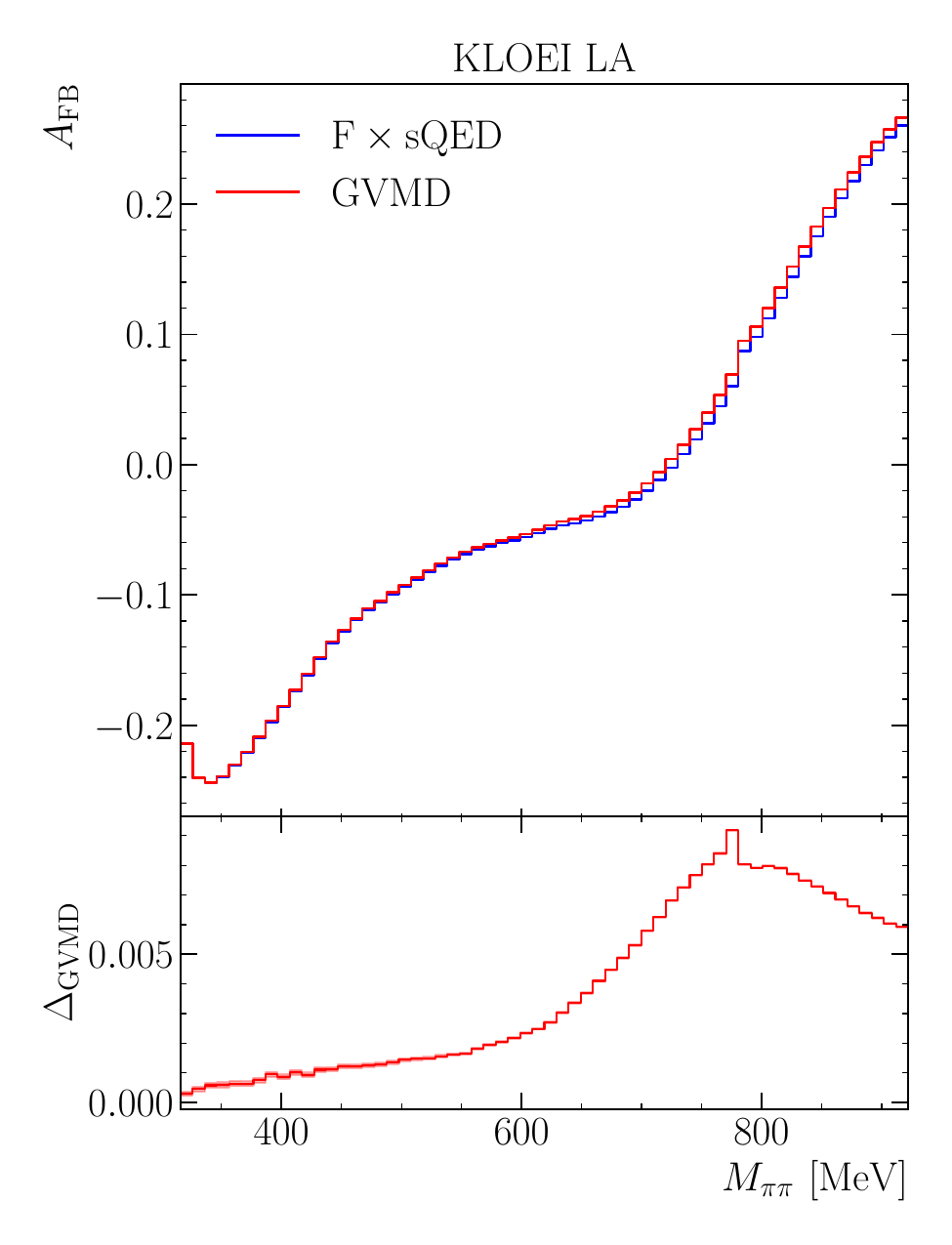} % Use \textwidth
\caption{The forward-backward asymmetry of the $e^+ e^- \to 
\pi^+ \pi^- \gamma$ process in the KLOEI LA scenario. The absolute difference in the lower panel is defined in Eq.~\eqref{eq:deltaa}.}
\label{fig:asymm}
\end{figure}

To model realistic event selection cuts used by radiative return experiments, we adopt the four scenarios described in~\cite{Aliberti:2024fpq} and also detailed in~\cite{Budassi:2026lmr}: KLOEI large-angle (LA) and KLOEII small-angle (SA) at c.m. energy of ${\sqrt{s}=1.02~{\rm GeV}}$, BESIII at ${\sqrt{s}=4~{\rm GeV}}$ and B at ${\sqrt{s}=10~{\rm GeV}}$.

We study the impact of the structure-dependent corrections, considering for all  scenarios the following differential cross sections:
\begin{equation}
    \frac{\rmd \sigma}{\rmd M_{\pi\pi}} \, ,
    \qquad \qquad \frac{\rmd \sigma}{\rmd \theta^+} \, ,
\end{equation}
where $\theta^+$ is the polar scattering angle of the $\pi^+$ particle, defined with respect to the direction of the incoming electron. The former observable is the quantity used in the experiments to extract the pion form factor from data, while the latter is relevant for acceptance studies.

Our results for the two-pion invariant mass and the angular distribution are shown in Fig.~\ref{fig:invmass} and 
Fig.~\ref{fig:angdist}, respectively. The relative contributions shown in the lower panels of the plots are defined as follows:
\begin{equation}
    {\rm K}^i_{\rm GVMD} \, = \, 
    \frac{{\rmd \sigma}^i_{\rm GVMD} - {\rmd \sigma}^i_{\rm F\times sQED}}{{\rmd \sigma}_{\rm F\times sQED}} \, ,
    \label{eq:Kfactor}
\end{equation}
where ${\rmd \sigma}$ is a short-hand notation that stands for the differential cross section as a function of $M_{\pi\pi}$ or $\theta^+$. In Eq.~\eqref{eq:Kfactor}, 
the superscript $i$ is given by $i = {\rm full}, (1\gamma^*), (2\gamma^*,{\rm ISR}), (2\gamma^*,{\rm FSR})$, where ``full'' denotes the complete correction. The separation of gauge-invariant contributions is performed at
the level of NLO amplitudes, which are interfered with the full
Born contribution.
% , whereas $a+b$, 
% $c$ and $d$ refer to the gauge-invariant topologies $a+b$, 
% $c$ and $d$ shown in Fig.~\ref{fig:topologies_gvmd}.

As can be seen from Fig.~\ref{fig:invmass}, the full GVMD corrections to the invariant mass around the $\rho$-peak are of the order of a few permille for all the experimental scenarios, from $\phi$- to $B$-factories. 
In the KLOEII SA scenario, the NLO corrections are dominated by the $(1\gamma^*)$ contribution due to the presence of a squared collinear logarithm $L^2$, where $L=\log\!\left(m_e^2/s\right)$. In the BESIII and B scenarios, the $(2\gamma^*)$ contributions are suppressed due to the small value of $F_\pi(s)$. Therefore, in these three scenarios the ${\rm K}_{\rm GMVD}^{\rm full}$ is essentially dominated by the $(1\gamma^*)$ subset.
The situation is different in the KLOEI LA scenario, where all topologies contribute with a similar size to the full correction, due to the different photon angular cut w.r.t. KLOEII SA.

% the size of $F_\pi(M_\rho^2)$ is not drastically different from $F_\pi(\sqrt{s})$ and the signal photon can become collinear to final state charged particle the structure-dependent contributions to ${\rmd \sigma} / {\rmd M_{\pi\pi}}$ are dominated by the $(1\gamma^*)$ topologies of Fig.~\ref{fig:topologies_gvmd}, i.e. the corrections to the Compton tensor and to the pion form factor. The only exception is the KLOEI LA setup, where all the topologies contribute.

The structure-dependent corrections to ${\rmd \sigma} / 
{\rmd \theta^+}$ are larger. As can be noticed from Fig.~\ref{fig:angdist}, in the 
$\theta^+$ angular distribution the GVMD predictions differ from ${\rm F\times sQED}$ at the percent level in all scenarios. In this case, the GVMD corrections are dominated by the Compton tensor topology of $(2\gamma^*,\rm ISR)$. This is due to the fact that the $(2\gamma^*,\rm ISR)$ diagrams contribute asymmetrically to $\rmd \sigma/\rmd \theta^+$ when interfered with the leading-order ISR, being enhanced by $|F_\pi(M_{\pi\pi}^2)|^2$ for $M_{\pi\pi}\simeq m_\rho$. 

To make contact with available experimental data 
and previous studies in the literature~\cite{Budassi:2026lmr,PetitRosas:2026iuq,WorkingGrouponRadiativeCorrections:2010bjp,Binner:1999bt}, we also investigate the forward-backward asymmetry $A_{\rm FB}$ in the KLOEI LA setup, which is defined as:
\begin{equation}
    A_{\rm FB}
    \left(M_{\pi\pi}\right)=\frac{\rmd \sigma_{\rm F}-\rmd \sigma_{\rm B}}{\rmd \sigma_{\rm F}+\rmd \sigma_{\rm B}} \, ,
    \label{eq:asymmetry}
\end{equation}
where
\begin{equation}
\begin{split}
    \rmd \sigma_{\rm F}&=\int_{0}^{1}\frac{\rmd \sigma}{\rmd M_{\pi\pi}\,\,\rmd \cos{\theta^+}} \rmd \cos{\theta^+} \, ,\\
    \rmd \sigma_{\rm B}&=\int_{-1}^{0}\frac{\rmd \sigma}{\rmd M_{\pi\pi}\,\,\rmd \cos{\theta^+}} \rmd \cos{\theta^+} \, .
\end{split}
\end{equation} 
Our results for the forward-backward asymmetry are shown in 
Fig.~\ref{fig:asymm}. The absolute difference shown in the lower panel is defined as:
\begin{equation}
    \Delta_{\rm GVMD} = A_{\rm FB}^{\rm GVMD} - 
    A_{\rm FB}^{\rm F\times sQED} \, .
    \label{eq:deltaa}
\end{equation}
From Fig.~\ref{fig:asymm}, one can see that the full GVMD correction amounts to about one percent close to the $\rho$-peak, but it is smaller below it.
In the ${e^+e^- \to \pi^+\pi^-}$ case, the forward-backward asymmetry is zero at LO and is induced at NLO by the IFI corrections, which are odd under the exchange of the pion momenta. Hence, as shown in~\cite{Ignatov:2022iou,Colangelo:2022lzg,Budassi:2024whw}, this observable is highly sensitive to the structure-dependent corrections.
In the ${e^+e^- \to \pi^+\pi^-\gamma}$ case, the forward-backward asymmetry is already present at LO. The GVMD corrections are an $\mathcal{O}(\alpha)$ effect and their absolute size is of the same order as in the $\pi^+\pi^-$ channel~\cite{Ignatov:2022iou,Colangelo:2022lzg,Budassi:2024whw}.

\begin{figure}[t] % [t] for top of page
\centering
\includegraphics[width=0.5\textwidth]{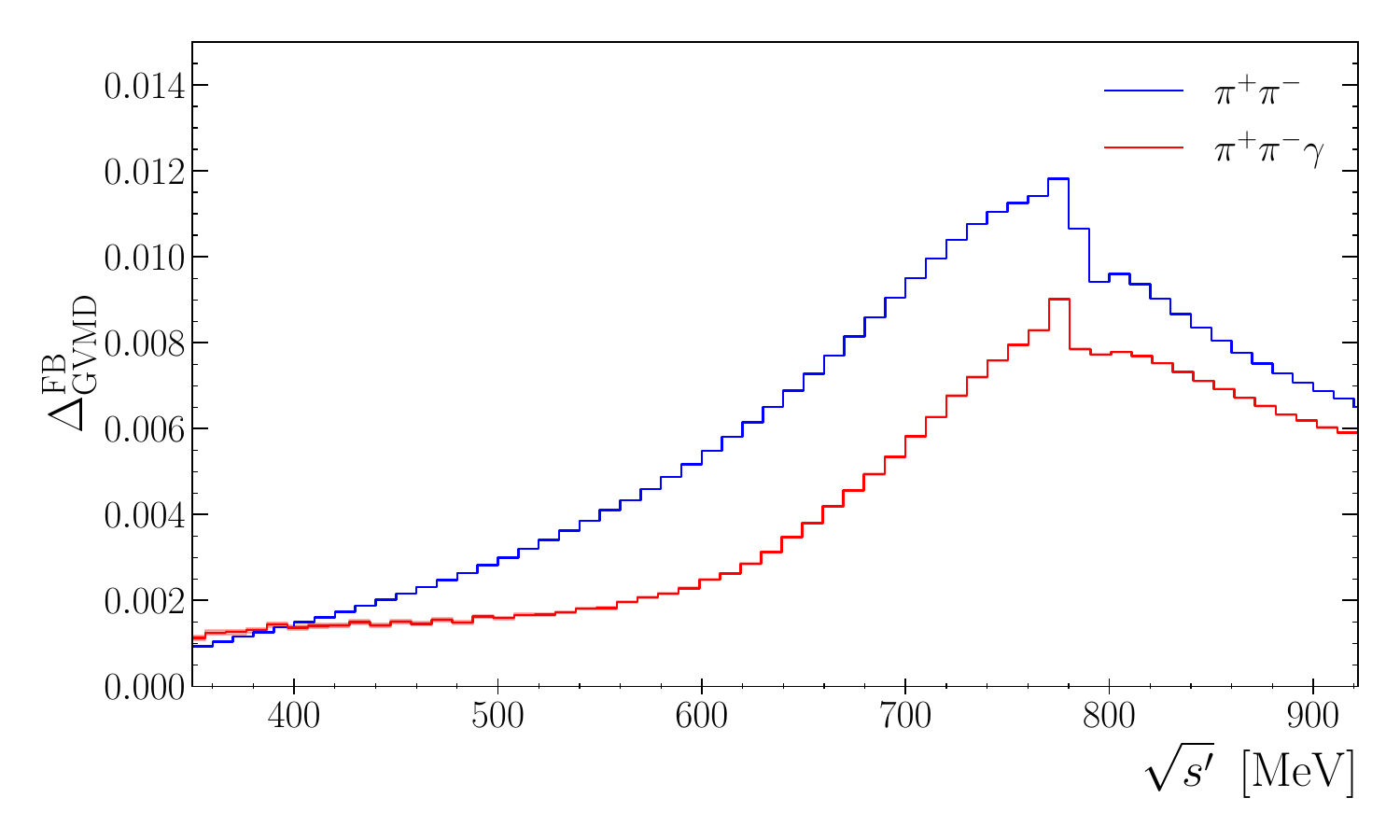} % Use \textwidth
\caption{The absolute GVMD corrections to the charge asymmetry ($\pi^+\pi^-$) in the CMD-3 setup of ~\cite{Budassi:2024whw} and to the FB asymmetry of the present study ($\pi^+\pi^-\gamma$). The quantity $\sqrt{s'}$ stands for $\sqrt{s}$ for the energy scan and $M_{\pi\pi}$ for radiative return. $\Delta_{\rm GVMD}^{\rm FB}$ is the absolute GVMD correction, namely $\Delta_{\rm GVMD}^{\rm FB}=A_{\rm GVMD}^{\rm FB}-A_{\rm F \times sQED}^{\rm FB}$.}
\label{fig:gvmdasymm}
\end{figure}

In both cases, the correction exhibits a resonant enhancement in the vicinity of the $\rho$ peak, as illustrated in Fig.~\ref{fig:gvmdasymm}. For a meaningful comparison 
between $A_{\rm FB}$ in $e^+ e^- \to \pi^+ \pi^- \gamma$ and 
in $e^+ e^- \to \pi^+ \pi^-$, it should be noted that the charge asymmetry in the energy scan process is defined in terms of $\theta_{\rm avg}=(\pi+\theta_- - \theta_+)/2$, whereas in the radiative return case the relevant observable is the 
forward-backward asymmetry as a function of $\theta_+$. Consequently, the sign of the GVMD correction obtained in the $2\to2$ process has been reversed in Fig.~\ref{fig:gvmdasymm}.
% On the other hand, in the ${e^+e^- \to \pi^+\pi^-\gamma}$ case, the forward-backward asymmetry is already present at LO and, therefore, the structure-dependent NLO contribution is only a $\mathcal{O}(\alpha)$ correction on top of it.
\section{Consistency checks}
\label{sec:cons_checks}

In this section, we describe further validation tests of our  results by making use of the GVMD corrections computed in~\cite{Budassi:2024whw} for the process ${e^+e^- \to \pi^+\pi^-}$ at NLOPS.

In the soft-photon approximation, where the emitted photon with momentum $k$ is soft, 
the QED virtual corrections to a generic radiative process involving $n+1$ photons 
can be written as the product of a universal eikonal factor and the virtual correction 
to the corresponding process with $n$ photons.  
This factorisation property also holds within the GVMD approach, making it possible to perform 
a direct comparison between the predictions obtained from the $2 \to 2$ code of~\cite{Budassi:2024whw} with NLOPS accuracy, expanded up to 
$\mathcal{O}(\alpha)$ with respect to the leading-order radiative process, and those derived 
from the present implementation, where the exact NLO result for the radiative process is fully included. These will be denoted, respectively, by the subscripts $\pi^+\pi^-$ and $\pi^+\pi^-\gamma$.
\begin{figure}[h]
    \centering
    \includegraphics[width=0.5\textwidth]{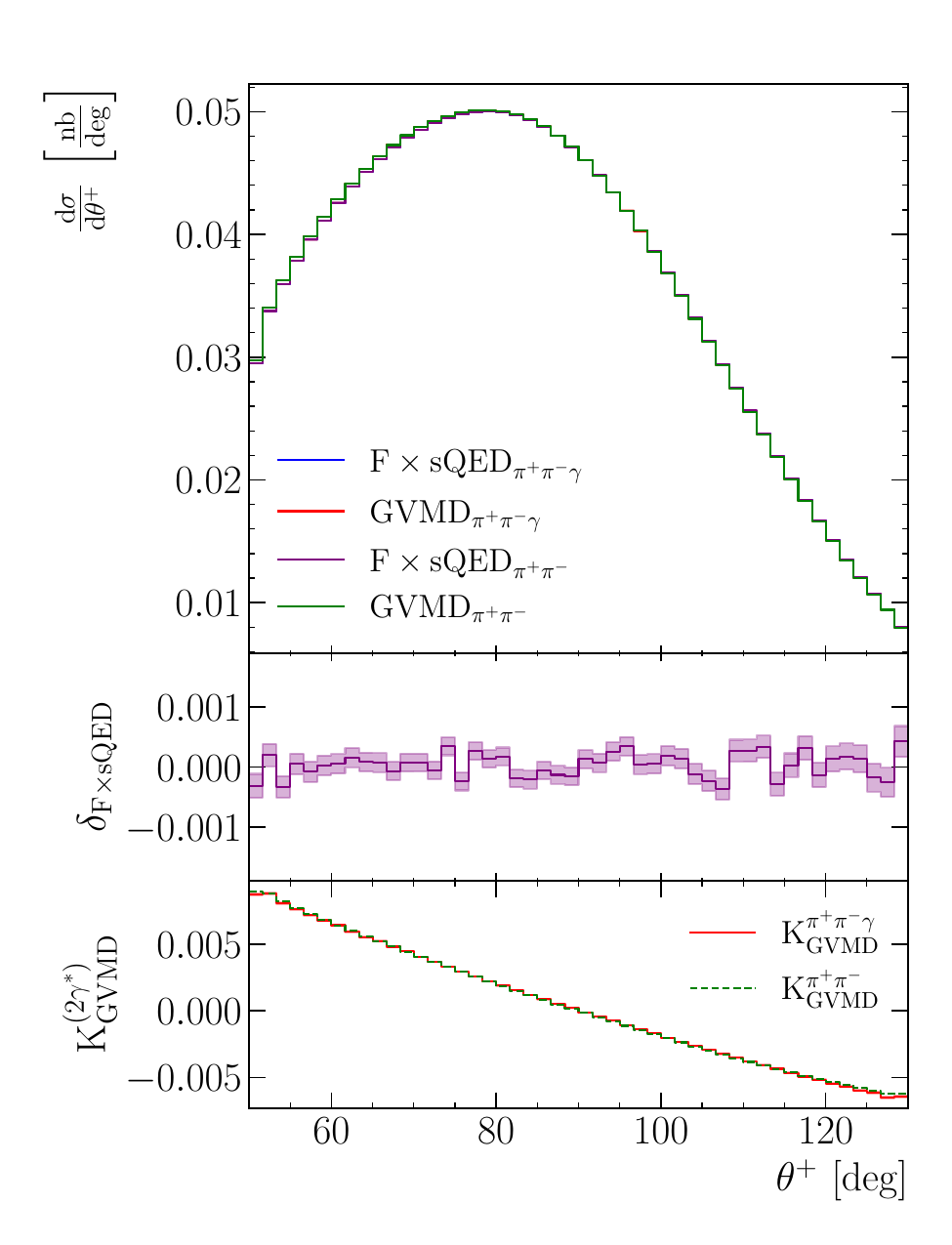}
    \caption{Differential cross section for the process $e^+e^-\to\pi^+\pi^-\gamma$ as a function of $\theta^+$ in the setup defined in Eq.~\eqref{eq:setup_soft}. In the middle panel, we show the relative difference as defined in Eq.~\eqref{eq:deltafsqed}, while the lower panel shows the ${\rm K}^{(2\gamma^*)}_{\rm GVMD}$ for the two different formulations discussed in the text. }
    \label{kloesa_thav}
\end{figure}

Thus, for event selections in which the soft-photon limit gives a good approximation of the full matrix element, 
we can expect that the quantity ${\rm K}_{\rm GVMD}$ obtained with the $2 \to 3$ exact NLO calculation 
is well reproduced by the $2 \to 2$ implementation with a tagged real photon. 
Specifically, we consider the following setup:
\begin{equation}
    \begin{split}
        &\sqrt{s}=1.02\,{\rm GeV} \, ,\\
        &50^\circ\le\theta_{\pm}\le 130^\circ \, ,\\
        &50^\circ\le\theta_{\gamma}\le 130^\circ\,,\;1\, {\rm MeV}\le E_\gamma\le 20 \,{\rm MeV}\,.
    \end{split}
    \label{eq:setup_soft}
\end{equation}

As discussed in~\cite{Budassi:2024whw} and corroborated by the results of this work, 
angular distributions are particularly sensitive to GVMD corrections, 
with the dominant contribution originating from the $(2\gamma^*)$ diagrams. For consistency, we compare the $2\to2$ and $2\to3$ implementations by restricting both of them to the contributions arising from these diagrams.

In Fig.~\ref{kloesa_thav}, we present the differential cross section as a function of $\theta^+$ in the scenario of Eq.~\eqref{eq:setup_soft}.
As can be seen from the lower panels, where 
\begin{equation}
    \delta_{\rm F\times sQED}=\frac{\rmd\sigma_{\rm F\times sQED}^{\pi^+\pi^-}-\rmd\sigma_{\rm F\times sQED}^{\pi^+\pi^-\gamma}}{\rmd\sigma_{\rm F\times sQED}^{\pi^+\pi^-\gamma}}\,\,,
    \label{eq:deltafsqed}
\end{equation}
the two factorised results agree at the $10^{-4}$~level. The ${\rm K}^{(2\gamma^*)}_{\rm GVMD}$ values computed using the two implementations are also in excellent agreement, providing an important internal check of the GVMD implementation of the
$\pi^+ \pi^- \gamma$ channel.

\begin{figure}[h]
        \centering \includegraphics[width=0.5\textwidth]{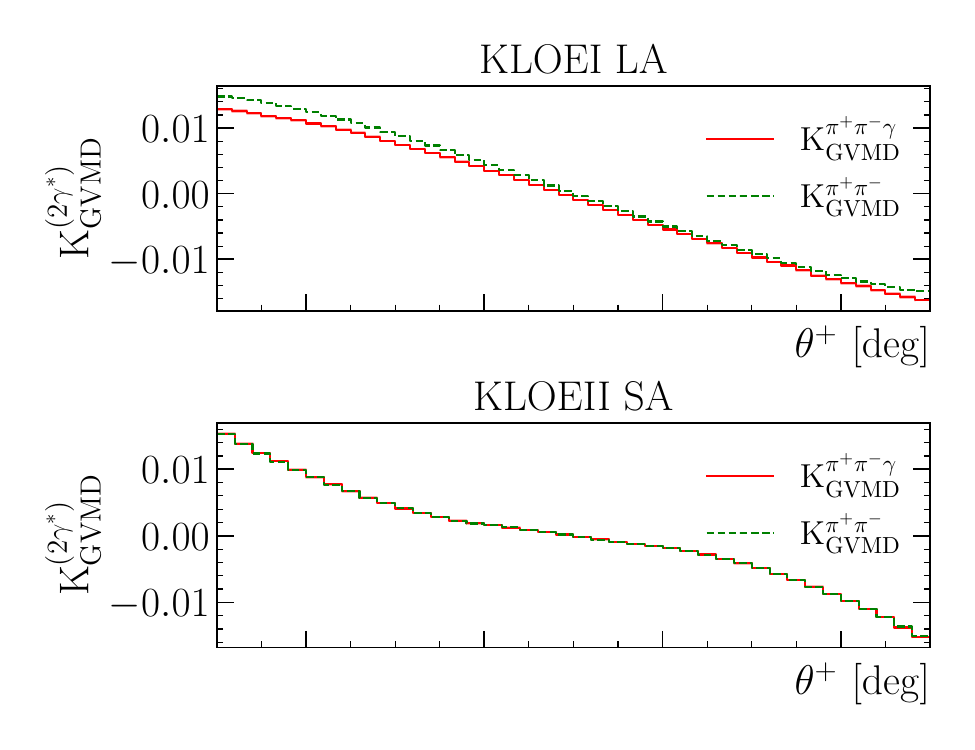}
    \caption{The same as the lower panel of Fig.~\ref{kloesa_thav} for the KLOEI LA and KLOEII SA scenarios.}
    \label{kloe_th+kgvmd}
\end{figure}

The same test can be extended to more realistic event selections. As an example, in Fig.~\ref{kloe_th+kgvmd}, we show for the KLOEI LA and KLOEII SA setups the behaviour of ${\rm K}^{(2\gamma^*)}_{\rm GVMD}$ as obtained within the two implementations. The agreement observed in the soft-photon limit is also confirmed in the KLOEII SA scenario, while the difference is at the permille level in the KLOEI LA setup.
\section{Conclusions}
\label{sec:conclusion}

We have computed the corrections due to the non-perturbative structure of the pion to the process $e^+ e^- \to \pi^+ \pi^- \gamma$, which is of primary importance for radiative return experiments at flavour factories. We have employed a GVMD approach to improve the commonly adopted F$\times$sQED recipe, including
the pion form factor in FSR and IFI one-loop integrals. We have implemented our calculation in the \textsc{BabaYaga@NLO} event generator to estimate the uncertainty in the modelling of the interaction of pions to photons in realistic experimental scenarios.

We have compared the GVMD predictions with the F$\times$sQED results for some observables of interest in radiative return measurements, considering experimental setups from $\phi$- to $B$-factories.
%From our study, we can conclude that the structure-dependent corrections are largely dominated by the IFI contributions, with a quite small impact from FSR. This is in agreement with what previously observed in the energy scan process $e^+ e^- \to \pi^+ \pi^-$.
We have observed that the invariant mass distribution receives GVMD corrections at the permille level in all experimental scenarios. 
On the other hand, the observables that are sensitive to the angular variables, such as 
the differential distribution of the scattering angle and the forward-backward asymmetry, show effects at the percent level. 
For the forward-backward asymmetry, the GVMD corrections exhibit a resonant enhancement around the $\rho$ peak, whose size is of the same order of magnitude as the one observed in the charge asymmetry of the energy scan process $e^+ e^- \to \pi^+ \pi^-$.

From our study, we can conclude that the structure-dependent corrections to $e^+ e^- \to \pi^+ \pi^- \gamma$ are, in general, not negligible for measurements of the pion form factor aiming at a sub-percent precision. At this level of precision, it is therefore mandatory to combine higher-order photon emissions~\cite{Budassi:2024whw} with structure-dependent corrections, being in general of the same order of magnitude.

During the completion of the present work, an independent computation of the structure-dependent corrections to 
$e^+ e^- \to \pi^+ \pi^- \gamma$ was performed in~\cite{PetitRosas:2026iuq} and interfaced to the \textsc{Phokhara} code. A comparison between our results and 
those of~\cite{PetitRosas:2026iuq} is left to future work.

As a next step, we are interested in improving our FSR model by including radiative meson decays, which play an important role around the $\phi$-resonance. We also plan to compute the one-loop corrections to $e^+ e^- \to \pi^+ \pi^- \gamma$ using the dispersive approach, with the aim of comparing the GVMD and dispersive predictions. Following the results already obtained 
in~\cite{Budassi:2024whw,Budassi:2026lmr}, the final goal is to arrive to a version of the \textsc{BabaYaga@NLO} generator including NLO structure-dependent corrections in two independent formalisms matched to a PS for the simulation of multi-photon emission for both $e^+ e^- \to \pi^+ \pi^-$ and $e^+ e^- \to \pi^+ \pi^- \gamma$. This version will be a valuable tool for new extractions of the pion form factor in both energy scan and radiative return experiments.

\section*{Acknowledgements} 

We wish to thank Mauro
Chiesa for useful discussions on automatic loop calculations and Pau Petit Ros{\`a}s for effective correspondence. 
FPU is grateful to the Instituto de Fisica Teorica
UAM/CSIC for hospitality and to the Erasmus+Traineeship program for partial support. 

\vskip 12pt\noindent
{\bf Data availability statement}
\vskip 8pt
The 
amplitudes derived in present work for the computation of 
structure-dependent corrections will be made publicly available as part of a future public release of the \textsc{BabaYaga@NLO}
generator through the project repository
\begin{center}
\href{https://github.com/cm-cc/BabaYagaNLO}{github.com/cm-cc/BabaYagaNLO }.
\end{center}

 \bibliographystyle{elsarticle-num} 
 \bibliography{GVMD}

\end{document}